\documentstyle[floats,aps]{revtex}
\begin{document}
\renewcommand{\thefootnote}{\fnsymbol{footnote}}
\draft
\title{\large\bf 
  Integrable  open-boundary conditions for the $q$-deformed supersymmetric
  $U$ model of strongly correlated electrons}

\author{Anthony J.  Bracken,  Xiang-Yu Ge \footnote
{E-mail: xg@maths.uq.edu.au},   Yao-Zhong Zhang 
     \footnote {Queen Elizabeth II Fellow.
                                   E-mail: yzz@maths.uq.edu.au}
             and 
        Huan-Qiang Zhou \footnote {On leave of absence from Dept of
	         Physics, Chongqing University, Chongqing 630044, China.
                 E-mail: hqzhou@cqu.edu.cn}} 

\address{Department of Mathematics,University of Queensland,
		     Brisbane, Qld 4072, Australia}

\maketitle

\vspace{10pt}

\begin{abstract}
A general graded reflection equation algebra is proposed and the 
corresponding boundary quantum inverse scattering method is formulated.
The formalism is applicable to all boundary lattice systems where an
invertible R-matrix exists. As an application, the integrable 
open-boundary conditions for the $q$-deformed supersymmetric
$U$ model of strongly correlated electrons are investigated.
The diagonal boundary K-matrices are found and a class of integrable
boundary terms are determined. The boundary system is solved
by means of the coordinate space Bethe ansatz technique and the Bethe ansatz
equations are derived. As a sideline, it is shown that all R-matrices
associated with a quantum affine superalgebra enjoy the crossing-unitarity
property.
\end{abstract}


%************************** Text Begins here ******************************

%  Greek letters

\def\a{\alpha}
\def\b{\beta}
\def\d{\delta}
\def\e{\epsilon}
\def\g{\gamma}
\def\k{\kappa}
\def\l{\lambda}
\def\o{\omega}
\def\t{\theta}
\def\s{\sigma}
\def\D{\Delta}
\def\L{\Lambda}

% Shorthands for \begin{equation} and the like

\def\beq{\begin{equation}}
\def\eeq{\end{equation}}
\def\bea{\begin{eqnarray}}
\def\eea{\end{eqnarray}}
\def\ba{\begin{array}}
\def\ea{\end{array}}
\def\no{\nonumber}
\def\le{\langle}
\def\re{\rangle}
\def\lt{\left}
\def\rt{\right}

\newcommand{\sect}[1]{\setcounter{equation}{0}\section{#1}}
\renewcommand{\theequation}{\thesection.\arabic{equation}}
\newcommand{\reff}[1]{eq.~(\ref{#1})}

%\newpage
\vskip.3in

\sect{Introduction\label{int}}

In the last decade, much attention has been paid to the study of
strongly correlated electron systems. In particular, the discovery
of high-$T_c$ superconductivity has greatly stimulated investigations
on various electron lattice models in one dimension (1D), which
are exactly solvable by means of the coordinate Bethe ansatz method
or the quantum inverse scattering method (QISM). The most known and
studied integrable electron systems are perhaps the Hubbard and the 
supersymmetric $t$-$J$ models. Other integrable correlated electron
theories of interest include, for instance,
the so-called extended Hubbard model \cite{Ess92/3},
the Hubbard-like model \cite{Bar91}, the supersymmetric $U$ model
\cite{Bra95} and its eight-state version \cite{Gou97}, and the
$q$-deformed supersymmetric $U$ model \cite{Bar95,Gou96}. These models
have been extensively investigated in the literature (c.f.:
\cite{Ess94,Zho96,Bed95}).

On the other hand, one of the recent developments in the theory of
completely integrable lattice models is Sklyanin's work \cite{Skl88}
on the boundary QISM, 
which may be used to describe integrable systems on a finite interval with
independent boundary conditions on each end. The important ingredient in this
boundary QISM is the new algebraic structure,
the reflection equation (RE)
algebra. The solutions to the REs are called boundary K-matrices which
in turn give rise to  boundary conditions
compatible with the bulk model integrability.

In Sklyanin's formulation on the boundary QISM, the R-matrices are
assumed to enjoy $P$-, $T$- and crossing-symmetry. Sklyanin's
results are generalized
by Mezincescu et al \cite{Mez91} to treat cases where, instead of having the
separate $P$- and $T$-symmetry, the R-matrices satisfy, in addition to
the crossing-symmetry, the less restrictive
condition of the combined $PT$ invariance. It is noted by de Vega et al
\cite{deV93} that the formalism of Mezincescu et al is actually also
applicable even if the R-matrices only have the so-called 
crossing-unitarity, a weaker version of the crossing-symmetry.

Generalizations to the following two cases seem very interesting: (i) the
graded or supersymmetric case, and (ii) the case where the R-matrices
do not obey any constraint conditions (except the
unitarity condition). Many attempts have been made in the literature
concerning the extension (i). However, very few authors treat
the grading properly from the beginning to the end. Therefore,
in the eyes of the present authors most known
results in the literature are not very satisfactory and a fully graded
or supersymmetric formalism is desirable.
Concerning (ii), two important examples are the R-matrices \cite{Zho96}
corresponding to the models of two coupled and three
coupled $XY$ spin chains introduced in \cite{Bar91}, to which  
the formalism developed in \cite{Skl88,Mez91} does not apply.

In section II of this paper, we shall fulfill the two aims of
extensions. More specifically, we shall formulate a fully supersymmetric
boundary QISM which is applicable to 
any cases where an invertible R-matrix exists. We
introduce a very general graded
RE algebra and show that this algebra indeed leads to a commuting
family of the boundary transfer matrices. Throughout the procedure,
no spectral parameter multiplicativity/additivity
of the graded quantum Yang-Baxter equation (QYBE) has been assumed and
no constraint conditions on the R-matrices (except the
always satisfied unitarity property) been imposed. 
Our formalism is a supersymmetric
generalization of that developed in \cite{Zhou96,Bra97} for
the bosonic (or non-supersymmetric) case, where
open-boundary integrable
conditions for the models of two coupled and three coupled 
one-dimensional $XY$ spin chains have been constructed, respectively. 
We then use, in section III, our results  to study integrable
open-boundary conditions
for the $q$-deformed supersymmetric $U$ model \cite{Bar95,Gou96} of strongly
correlated electrons. The quantum integrability of the model in the
bulk has been established in \cite{Gou96}
by embedding the bulk model Hamiltonian into a one-parameter family
of commuting transfer matices formed from a $U_q[gl(2|1)]$ invariant
R-matrix.  
We solve the graded REs for the diagonal boundary K-matrices and  determine
the open-boundary integrable model. 
The boundary model Hamiltonian is shown
to be related to the second derivative of the boundary transfer matrix.
In section IV, we solve this boundary model by the coordinate 
space Bethe ansatz method and derive the Bethe ansatz equations.
In the Appendix, we show that all R-matrices associated with
a quantum affine superalgebra possess the crossing-unitarity.

\sect{Graded Reflection Equations and Transfer Matrix: 
      General Formulation\label{for}}

In this section, we establish a very general RE algebra. We shall not
assume spectral-parameter multiplicativity/additivity of the graded QYBE,
although for the special example we 
shall discuss in section III 
the spectral parameters are multiplicative/additive. 
Throughtout the procedure, no
constraint conditions have been assumed on the R-matrices.

To begin with, let
$V$ be a finite-dimensional ${\bf Z}_2$ graded linear superspace. Let
$R\in End(V\otimes V)$ be a solution to the graded QYBE
\beq
R_{12}(u_1,u_2)R_{13}(u_1,u_3)R_{23}(u_2,u_3)=R_{23}(u_2,u_3)
R_{13}(u_1,u_3)R_{12}(u_1,u_2).
\eeq
Here $R_{jk}(u)$ denotes the matrix on $V\otimes V\otimes V$ acting on the
$j$-th and $k$-th superspaces and as an identity on the remaining 
superspace. The
variables $u_1,~u_2$ and $u_3$ are  spectral parameters. The 
tensor product
should be understood in the graded sense, that is the
multiplication rule for any homogeneous elements $x,y,x',y'\in V$ is
given by
\beq
(x\otimes y) (x'\otimes y')=(-1)^{[y][x']}\;(xx'\otimes yy')
\eeq
where $[x]$ stands for the grading of element $x$: i.e. $[x]=0$ if $x$
is even (or bosonic) and $[x]=1$ if $x$ is odd (or fermionic). Let $P$
be the graded permutation operator in $V\otimes V$. Then
$P(x\otimes y)=(-1)^{[x][y]}y\otimes x$ for all homogeneous $x,y\in V$ and
$R_{21}(u)=P_{12}R_{12}(u)P_{12}$. 
We form the monodromy matrix
$T(u)$ for an $L$-site lattice  chain 
\beq
T(u)=L_{0L}(u)\cdots L_{01}(u),
\eeq
where $L_{0j}(u) \equiv R_{0j}(u,0)$, where subscript $0$ labels the auxiliary
superspace $V$. Indeed, one may
show that $T(u)$ generates  a representation of the graded 
quantum Yang-Baxter algebra,
\beq
R_{12}(u_1,u_2)\stackrel {1}{T}(u_1) \stackrel {2}{T}(u_2)
  =\stackrel {2}{T}(u_2)
  \stackrel {1}{T}(u_1) R_{12}(u_1,u_2),\label{yb-alg}
\eeq
where $\stackrel {1}{X} \equiv  X \otimes 1$ and
$\stackrel {2}{X} \equiv  1 \otimes X$,
for any matrix $ X \in End(V) $. 

In order to construct integrable electronic models with open boundary
conditions, we need to introduce an appropiate graded RE algebra. 
We introduce two associative superalgebras ${\cal T}_-$ and ${\cal T}_+$
defined by the R-matrices and the relations
\bea
R_{12}(u_1,u_2)\stackrel {1} {\cal T}_-(u_1) R_{21}(u_2,-u_1)
\stackrel {2}{\cal T}_-(u_2)&=& 
\stackrel {2}{\cal T}_-(u_2) R_{12}(u_1,-u_2)
\stackrel {1}{\cal T}_-(u_1) R_{21}(-u_2,-u_1),\no\\
R_{21}^{st_1 \; ist_2}(u_2,u_1)\stackrel {1}{{\cal T}^{st_1}_+}(u_1) 
{\tilde R}_{12}(-u_1,u_2)
\stackrel {2}{{\cal T}^{ist_2}_+}(u_2)& = &
\stackrel {2}{{\cal T}^{ist_2}_+}(u_2) {\tilde{\tilde R}}_{21}(-u_2,u_1)
\stackrel {1}{{\cal T}^{st_1}_+}(u_1) R_{12}^{st_1\; ist_2}(-u_1,-u_2),
\label{re-alg}    
\eea
where we have defined  new objects ${\tilde R}$ 
and  ${\tilde {\tilde R}}$ 
through the relations
\bea
{\tilde R} _{12}^{st_2} (-u_1,u_2) R^{st_1}_{21}(u_2,-u_1)&=&1,\no\\
{\tilde {\tilde {R}}} _{21}^{ist_1} (-u_2,u_1) R^{ist_2}_{12}(u_1,-u_2)&=&1,
\label{cross}
\eea
and, as defined in the Appendix,
$st_i$ stands for the supertransposition taken in the $i$-th space,
whereas $ist_i$  is the operation inverse to $st_i$. As will become
clear below, one of the important steps towards formulating a
correct formalism for the graded or supersymmetric case is to introduce,
in the second RE in (\ref{re-alg}) below,
the inverse operation of the supertransposition. The introduction of
this inverse operation is essential
because, applying the same supertransposition operation twice does not 
in general give an identity operation. For instance, applying
$st_i,~i=1,2$ ,twice to the R-matrix $R(u)$ does not yield $R(u)$:
\beq
\{R(u)^{st_i}\}^{st_i}\neq R(u),~~~i=1,2.
\eeq
In all cases, the quantum R-matrices possess the unitarity \footnote{
One can always normalize the R-matrices so that the right hand side of 
(\ref{unitarity}) equals to 1},
\beq
 R_{12}(u_1,u_2)R_{21}(u_2,u_1) = 1.\label{unitarity}
\eeq

We now show that the second RE in
(\ref{re-alg}) is indeed the correct ``conjugation" to the first one, so
that the boundary transfer matrices defined as usual constitute a  commuting
familty. Following  Sklyanin's arguments \cite{Skl88},
one defines the boundary transfer matrix $t(u)$ as
\beq
t(u) = str ({\cal T}_+(u){\cal T}_-(u)),
\eeq
where $str$ denotes the supertrace taken over the auxiliary superspace
$V$. Then it can be shown that
\beq
[t(u_1),t(u_2)] = 0.
\eeq
The proof is elementary. We nevertheless present the details. By means
of the commutativity of operators in ${\cal T}_+$ and ${\cal T}_-$,
\bea
t_1(u_1)t_2(u_2)&=&str_1\{\stackrel{1}{{\cal T}_+}(u_1)\stackrel{1}{{\cal T}_-}
   (u_1)\}str_2\{\stackrel{2}{{\cal T}_+}(u_2)\stackrel{2}{{\cal T}_-}(u_2)
   \}\no\\
&=&str_1\{\stackrel{1}{{\cal T}_+^{st_1}}(u_1)\stackrel{1}{{\cal T}_-^{st_1}}
   (u_1)\}str_2\{\stackrel{2}{{\cal T}_+}(u_2)\stackrel{2}{{\cal T}_-}(u_2)
   \}\no\\
&=&str_{12}\{\stackrel{1}{{\cal T}_+^{st_1}}(u_1)\stackrel{2}{{\cal T}_+}(u_2)
   \stackrel{1}{{\cal T}_-^{st_1}}
   (u_1)\stackrel{2}{{\cal T}_-}(u_2)\}=\cdots~.\no
\eea
Inserting a variant of the first expression in (\ref{cross}) for ``1" into the 
supertrace,
$$
\cdots=str_{12}\{\stackrel{1}{{\cal T}_+^{st_1}}(u_1)\stackrel{2}{{\cal
   T}_+}(u_2)
   \tilde{R}^{st_2}_{12}(-u_1,u_2)R^{st_1}_{21}(u_2,-u_1)
   \stackrel{1}{{\cal T}_-^{st_1}}(u_1)
   \stackrel{2}{{\cal T}_-}(u_2)\}=\cdots\,,
$$
then applying the supertransposition,
\bea
\cdots&=&str_{12}\{\stackrel{1}{{\cal T}_+^{st_1}}(u_1)
   \tilde{R}_{12}(-u_1,u_2)
   \stackrel{2}{{\cal T}_+^{ist_2}}(u_2)\}^{st_2}
   \{\stackrel{1}{{\cal T}_-}(u_1)
   R_{21}(u_2,-u_1)
   \stackrel{2}{{\cal T}_-}(u_2)\}^{st_1}\no\\
&=&str_{12}\{\stackrel{1}{{\cal T}_+^{st_1}}(u_1)
   \tilde{R}_{12}(-u_1,u_2)
   \stackrel{2}{{\cal T}_+^{ist_2}}(u_2)\}^{ist_1 \; st_2}
   \{\stackrel{1}{{\cal T}_-}(u_1)
   R_{21}(u_2,-u_1)
   \stackrel{2}{{\cal T}_-}(u_2)\}\no\\
&=&\cdots,\no
\eea
and, finally, inserting the unitarity property into the supertrace one obtains
\bea
\cdots&=&str_{12}[\{\stackrel{1}{{\cal T}_+^{st_1}}(u_1)
   \tilde{R}_{12}(-u_1,u_2)
   \stackrel{2}{{\cal T}_+^{ist_2}}(u_2)\}^{ist_1 \; st_2}
   R_{21}(u_2,u_1)\no\\
& &~~~~~~\times   R_{12}(u_1,u_2)\{\stackrel{1}{{\cal T}_-}(u_1)
   R_{21}(u_2,-u_1)
   \stackrel{2}{{\cal T}_-}(u_2)\}]\no\\
&=&str_{12}[\{R_{21}^{st_1 \; ist_2}(u_2,u_1)
   \stackrel{1}{{\cal T}_+^{st_1}}(u_1)
   \tilde{R}_{12}(-u_1,u_2)
   \stackrel{2}{{\cal T}_+^{ist_2}}(u_2)\}^{ist_1 \; st_2}\no\\
& & ~~~~~~\times R_{12}(u_1,u_2)\stackrel{1}{{\cal T}_-}
   R_{21}(u_2,-u_1)
   (u_1)\stackrel{2}{{\cal T}_-}(u_2)]
   =\cdots~.\no
\eea
Applying the RE algebra (\ref{re-alg}),
\bea
\cdots&=&str_{12}[\{\stackrel{2}{{\cal T}_+^{ist_2}}(u_2)
   \tilde {\tilde{R}}_{21}(-u_2,u_1)
   \stackrel{1}{{\cal T}_+^{st_1}}(u_1)R_{12}^{st_1 \; ist_2}(-u_1,-u_2)\}
   ^{ist_1 \; st_2}\no\\
& &~~~~~~\times   \stackrel{2}{{\cal T}_-}(u_2)
   R_{12}(u_1,-u_2)
   \stackrel{1}{{\cal T}_-}(u_1)R_{21}(-u_2,-u_1)]\no\\
&=&str_{12}[\stackrel{2}{{\cal T}_+^{ist_2}}(u_2)
   \tilde {\tilde{R}}_{21}(-u_2,u_1)
   \stackrel{1}{{\cal T}_+^{st_1}}(u_1)R_{12}^{st_1 \; ist_2}(-u_1,-u_2)\no\\
& &~~~~~~\times   \{\stackrel{2}{{\cal T}_-}(u_2)
   R_{12}(u_1,-u_2)
   \stackrel{1}{{\cal T}_-}(u_1)R_{21}(-u_2,-u_1)\}^{st_1 \; ist _2}].\no
\eea
Repeating the whole chain of transformation in the reverse order and
keeping in mind the second expression in (\ref{cross}) for ``1", one
ends up with $t(u_2)t(u_1)$, as required.

One can obtain a class of realizations of the superalgebras ${\cal T}_+$  and
${\cal T}_-$  by choosing  ${\cal T}_{\pm}(u)$ to be the form
\beq
{\cal T}_-(u) = T_-(u) K_-(u) T^{-1}_-(-u),~~~~~~ 
{\cal T}_+(u) = K_+(u)\label{t-,t+} 
\eeq
with
\beq
T_-(u) = R_{0N}(u) \cdots R_{01}(u),
\eeq
where $N$ is any node between 1 and $L$ (including $1$ and $L$), and 
$K_{\pm}(u)$, called boundary K-matrices, 
are representations of  ${\cal T}_{\pm}  $ in a Grassmann algebra.
In this realization, the elements of the matrices $K_\pm(u)$ are all Grassmann
numbers. Note that the K-matrices $K_\pm(u)$ satisfy the same
relations as ${\cal T}_\pm(u)$, respectively. That is, the K-matrices
fulfill the following graded REs deduced from (\ref{re-alg})
\bea
R_{12}(u_1,u_2)\stackrel {1} {K}_-(u_1) R_{21}(u_2,-u_1)
\stackrel {2}{K}_-(u_2)&=& 
\stackrel {2}{K}_-(u_2) R_{12}(u_1,-u_2)
\stackrel {1}{K}_-(u_1) R_{21}(-u_2,-u_1),\no\\ 
R_{21}^{st_1 \; ist_2}(u_2,u_1)\stackrel {1}{K^{st_1}_+}(u_1) 
{\tilde R}_{12}(-u_1,u_2)
\stackrel {2}{K^{ist_2}_+}(u_2)& = &
\stackrel {2}{K^{ist_2}_+}(u_2) {\tilde{\tilde R}}_{21}(-u_2,u_1)
\stackrel {1}{K^{st_1}_+}(u_1) R_{12}^{st_1\; ist_2}(-u_1,-u_2).\label{REs}    
\eea

The REs (\ref{re-alg}) and their realizations (\ref{REs}) in the
Grassmann algebra are generalizations of those introduced 
in \cite{Zhou96,Zha97}. Note that no constraint conditions have been
imposed on the
R-matrices. Therefore our graded REs apply to any case 
where an invertible R-matrix exists.

Let us now consider the important ``special" case in which the
R-matrices are related to finite dimensional representations $\pi_V$ of
the quantum affine superalgebra $U_q[{\cal G}^{(k)}~(k=1,2)$ for 
generic $q$, where ${\cal G}$ is any simple Lie superalgebra.
In the Appendix, we show that all such R-matrices
enjoy the crossing-unitarity properties [see (\ref{cu-twisted}) or
(\ref{cu-twisted-bar})]. One can show that there exists a more general
realization of the superalgebra ${\cal T}_+$ 
\beq
{\cal T}^{st}_+(u) = T^{st}_+(u) K^{st}_+(u) 
  \lt(T^{-1}_+(-u)\rt)^{st},~~~~~~
T_+(u) = R_{0L}(u) \cdots R_{0,N+1}(u).
\eeq
In the sequel, without loss of generality, we shall choose
$N=L$ so that ${\cal T}_+(u)\equiv K_+(u)$.
As the spectral parameters in the QYBE are now
multiplicative/additivity,  the K-matrices fulfill  the REs of
the following form,
\bea
R_{12}(u_1-u_2)\stackrel {1} {K}_-(u_1) R_{21}(u_1+u_2)
\stackrel {2}{K}_-(u_2)&=& 
\stackrel {2}{K}_-(u_2) R_{12}(u_1+u_2)
\stackrel {1}{K}_-(u_1) R_{21}(u_1-u_2),\no\\ 
R_{12}(-u_1+u_2)\stackrel {1}{K_+}(u_1) 
{\tilde{\tilde{R}}_{21}^{ist_1\;st_2}(-u_1-u_2)
\stackrel {2}{K}_+}(u_2)& = &
\stackrel {2}{K_+}(u_2) {\tilde R}_{12}^{ist_1\;st_2}(-u_1-u_2)
\stackrel {1}{K_+}(u_1) R_{21}(-u_1+u_2),\label{REs-with-cu}    
\eea
where in deriving the second relation we have applied the operation 
$ist_1\;st_2$ on both side of the second equation of (\ref{REs}).

By (\ref{cross}),
\bea
\tilde{\tilde{R}}_{21}^{ist_1\;st_2}(-u_1-u_2)&=&(((R_{21}(-u_1-u_2)^{-1}
  )^{ist_2})^{-1})^{st_2},\no\\
{\tilde R}_{12}^{ist_1\;st_2}(-u_1-u_2)&=&(((R_{12}(-u_1-u_2)^{-1}
  )^{st_1})^{-1})^{ist_1}.
\eea
With the help of (\ref{cu-twisted}),
(\ref{ist-st}) and (\ref{rho}) (one should identify $z=q^u$ to convert the
multiplicative spectral parameter $z$ into the additive one $u$),
one can show that the second RE in (\ref{REs-with-cu}) becomes
\bea
&&R_{12}(-u_1+u_2)\stackrel {1}{K_+}(u_1)\stackrel{1}{M^{-1}} 
R_{21}(-u_1-u_2+\frac{2}{k}g)\stackrel{1}{M}
\stackrel {2}{K_+}(u_2)\no\\
&&~~~~~~~~~~~~~~~~~~~ =\stackrel{1}{M} 
\stackrel {2}{K_+}(u_2) R_{12}(-u_1-u_2+\frac{2}{k}g)\stackrel{1}{M^{-1}}
\stackrel {1}{K_+}(u_1) R_{21}(-u_1+u_2),\label{RE-with-cu}
\eea
where $M=\pi_V(q^{-2\rho})$ is the so-called crossing matrix and $g$ is
defined as in the Appendix. 
The RE (\ref{RE-with-cu}) coincides, in the case of $k=1$, with
the one used in \cite{Lin96} and is applicable to all cases whose
R-matrices are related to the quantum affine superalgebra
$U_q[{\cal G}^{(k)}]$ for generic $q$.

\sect{Integrable boundary K-matrices and $q$-deformed supersymmetric
      $U$ model with boundary terms\label{Boun}}

Let $c_{j,\s}$ and $c_{j,\s}^{\dagger}$ denote fermionic creation and
annihilation operators with spin $\s$ at
site $j$, which satisfy the anti-commutation relations given by
$\{c_{i,\s}^\dagger, c_{j,\tau}\}=\d_{ij}\d_{\s\tau}$, where 
$i,j=1,2,\cdots,L$ and $\s,\tau=\uparrow,\;\downarrow$. We consider the
$q$-deformed supersymmetric $U$ model with boundary terms 
described by the following Hamiltonian:
\beq
H=\sum _{j=1}^{L-1} H_{j,j+1}^Q + B_L +B_R,\label{h}\label{hamiltonian}
\eeq
where $H_{j,j+1}^Q$ is the bulk Hamiltonian density of the $q$-deformed
supersymmetric $U$ model \cite{Bar95,Gou96}
\bea
H_{j,j+1}^Q&=&-\sum _{\sigma}(c^{\dagger}_{j\sigma}c_{j+1\sigma}+h.c.)
  \exp(-\frac {1}{2}(\eta -\s \g)n_{j,-\sigma}-\frac {1}{2}
  (\eta + \s \g)n_{j+1,-\sigma})\no\\
& &+\frac {U}{2}(n_{j\uparrow}n_{j\downarrow}
  +n_{j+1\uparrow}n_{j+1\downarrow})\no\\
& &+t_p(c^{\dagger}_{j\uparrow}c^{\dagger}_{j\downarrow}c_{j+1\downarrow}
  c_{j+1\uparrow}+h.c.)+\mu\; n_{j}+\mu ^{-1}\;n_{j+1},\label{h-density}
\eea
and $B_L$ and $B_R$ are left and right boundary terms, respectively,
given by
\bea
B_L&=&-\frac {\mu -\mu ^{-1}}
  {2\sinh \frac {\g(2-\xi_-)}{2}}
\lt(\frac {\sinh \g}{\sinh \frac {\g \xi_-}{2}}
  n_{1\uparrow}n_{1\downarrow}
  -e^{-\g (1-\frac {\xi_-}{2})}n_1\rt),\no\\
B_R&=&-\frac {\mu -\mu ^{-1}}
  {2\sinh \frac {\g(2-\xi_+)}{2}}
\lt(\frac {\sinh \g}{\sinh \frac {\g \xi_+}{2}}
  n_{L\uparrow}n_{L\downarrow}
  -e^{\g (1-\frac {\xi_+}{2})}n_L\rt),
\eea
where $n_{j\s}$ is the density operator
$n_{j\s}=c_{j\s}^{\dagger}c_{j\s}$,
$n_j=n_{j\uparrow}+n_{j\downarrow}$ and 
\bea
t_p&=&\frac {U}{2} =
  \e [2e^{-\eta}(\cosh \eta -\cosh \g)]^{\frac {1}{2}},~~~~\e =\pm,\no\\
\mu& =&\sqrt{ e^{\g}\frac {\sinh (\eta -\g)/2}{\sinh (\eta +\g)/2}};
\eea
$\xi_{\pm}$ are some parameters describing the
boundary effects. It is interesting to note that the Hamiltonian
(\ref{hamiltonian}) 
becomes $U_q[gl(2|1)]$-invariant in the limits $\xi _- \rightarrow
-\infty,~~\xi _+ \rightarrow \infty.$

We shall establish quantum integrability for the system defined by the
Hamiltonian (\ref{hamiltonian}), by applying the general formalism
developed in the previous section. To this end, let
us first of all recall some basic results of the 
$q$-deformed supersymmetric $U$ model with the periodic boundary conditions.
In \cite{Gou96}, it was shown that the bulk Hamiltonian
(\ref{h-density}) of the model commutes with 
the bulk transfer matrix $\tau(u)$, which is the supertrace of the
monodromy matrix $T(u)$ with the local monodromy matrix $L_{0j}(u)=
R_{0j}(u)$. That is,
\beq
\tau(u)=str(T(u)),~~~~T(u) = R_{0L}(u)\cdots R_{01}(u), \label{matrix-t}
\eeq
where the quantum R-matrix $R(u) \equiv
P \check {R}(u)$, with 
\beq
\check { R}(u) = \frac {q^u-q^{2 \a}}{1-q^{u+2 \a}}
  \check{P}_1 +\check{P}_2 + \frac{1-q^{u+2 \a +2}}{q^u-q^{2 \a +2}}
  \check{P}_3,\label{r}
\eeq
where $\check{P}_i,~i=1,2,3$ are
the projection operators whose explicit formulae may be found in \cite{Gou96}.

In order to describe integrable systems with the boundary conditions different
from the periodic ones, we first solve the REs for the two boundary
K-matrices $K_{\pm}(u)$. 
For our purpose, we only look for solutions where $K_\pm(u)$ are
diagonal. After complicated algebraic manipulations, we find
\beq
K_-(u)=  \frac {1}{ \sinh \frac {\g \xi_-}{2}\sinh \frac {\g(\xi_--2)}{2}}\;
\left ( \begin {array}{cccc}
A_-(u)&0&0&0\\
0&B_-(u)&0&0\\
0&0&B_-(u)&0\\
0&0&0& C_-(u)
\end {array} \right ),\label{k-}
\eeq
where
\bea
A_-(u)&=&e^{\g u} \sinh  \frac {\g (\xi_-+u)}{2}\sinh \frac {\g (u-2+\xi_-)}{2},\no\\
B_-(u)&=& \sinh \frac {\g (\xi_--u)}{2}\sinh \frac {\g (u-2+\xi_-)}{2},\no\\
C_-(u)&=&e^{-\g u} \sinh \frac {\g (\xi_--u)}{2}\sinh \frac {\g (-u-2+\xi_-)}{2}.
\eea
As is shown in the Appendix, the R-matrix (\ref{r}) satisfies the
crossing-uniatriy condition. This 
implies that there is an isomorphism between the graded REs
for $K_+$ and $K_-$:
\beq
K_+(u)=M K_-(-u+1), 
\eeq
where $M$ is given by (up to an overall factor)
\beq
M=\left ( \begin {array} {cccc}
1&0&0&0\\
0&1&0&0\\
0&0& e^{2\g}&0\\
0&0&0&e^{2\g} 
\end {array} \right ).\label{m}
\eeq
Therefore we may choose the boundary K-matrix $K_+(u)$ as
\beq
K_+(u)=\left ( \begin {array} {cccc}
A_+(u)&0&0&0\\
0&B_+(u)&0&0\\
0&0& C_+(u)&0\\
0&0&0& D_+(u)
\end {array} \right )\label{k+}
\eeq
with
\bea
A_+(u)&=& e^{-\g(u-1)} \sinh \frac {\g(2-2\alpha-\xi_+-u)}{2}
\sinh \frac{\g(2\alpha+\xi_++u)}{2},\no\\
B_+(u)&=&e^{-2\g} \sinh \frac {\g(-2\alpha-\xi_++u)}{2}
\sinh \frac {\g(2\alpha+\xi_++u)}{2},\no\\
C_+(u)&=& \sinh \frac {\g(-2\alpha-\xi_++u)}{2}
\sinh \frac {\g(2\alpha+\xi_++u)}{2},\no\\
D_+(u)&=&e^{\g(u-1)}\sinh \frac {\g(-2\alpha-\xi_++u)}{2}
\sinh \frac {\g(2+2\alpha+\xi_+-u)}{2}.
\eea
Indeed,we have checked that this $K_+$ matrix constitutes a solution
to the graded REs (\ref{RE-with-cu}).

To  show that the Hamiltonian (\ref{hamiltonian}) can be embedded 
into the boundary transfer matrix
$t(u)$ constructed in section II is
an involved algebraic manipulation. This is because the supertrace of $K_+(0)$
is equal to zero. So at best we can only  expect 
that the Hamiltonian (\ref{hamiltonian})
appears as the second derivative of the boundary transfer matrix with respect
to the spectral parameter $u$, at $u=0$. 

Let us expand the
local monodromy matrix $L_{0j}(u)$ up to the  second order in the
spectral parameter $u$,
\beq
L_{0j}(u)= (1+H_{j0}u +\frac {1}{2!}B_{j0}u^2+ 
+\cdots)L_{0j}(0).
\eeq
Substituting this expression into the boundary 
transfer matrix $t(u)$, and after a
lengthy but straightforward algebraic calculation, one finds
\beq
t(u) =  C_1 u + C_2 (H + const.) u^2 + \cdot\cdot\cdot,
\eeq
where $C_i (i = 1,2,\cdots)$ are some scalar functions of the boundary
constant $\xi _+$. Then it can be shown that up to some additive 
constants the Hamiltonian (\ref{hamiltonian}) is related to 
the second derivative of the boundary transfer matrix, 
\bea
H&=&-\frac{q^{\a +1}-q^{-\a -1}}{\ln q}H^R,\no\\
H^R&=&\frac {t'' (0)}{4(V+2W)}=
  \sum _{j=1}^{L-1} H^R_{j,j+1} + \frac {1}{2} \stackrel {1}{K'}_-(0)
+\frac {1}{2(V+2W)}\lt[str_0\lt(\stackrel {0}{K}_+(0)G_{L0}\rt)\rt.\no\\
& &\lt.+2\,str_0\lt(\stackrel {0}{K'}_+(0)H_{L0}^R\rt)+
  str_0\lt(\stackrel {0}{K}_+(0)\lt(H^R_{L0}\rt)^2\rt)\rt],\label{derived-h}
\eea
where  
\bea
V&=&str_0 K'_+(0),
~~W=str_0 \lt(\stackrel {0}{K}_+(0) H_{L0}^R\rt),\no\\
H^R_{i,j}&=&P_{i,j}R'_{i,j}(0),
~~G_{i,j}=P_{i,j}R''_{i,j}(0),
\eea
if we make the following identifications:
\beq
q=e^\g,~~~~~~~~ \frac{q^{\a+1}-q^{-\a-1}}{q^\a-q^{-\a}}=e^{-\eta}.
\eeq
Thus, we have shown that the  Hamiltonian (\ref{hamiltonian}) of the 
$q$-deformed supersymmetric $U$ model with the boundary
terms $B_L$ and $B_R$
is related to a class of commuting transfer matrices.
As a result, the system has an infinite number of higher conserved currents
which are involutive with each other, and therefore
the system under study is completely integrable.

\sect{The Bethe ansatz equations \label{bethe}}

Having established the quantum integrability of the model,
let us  in this section  solve it by using
the coordinate space Bethe ansatz method. Following \cite{Asa96,Zha97,Bra97},
we assume that the eigenfunction of Hamiltonian (\ref{hamiltonian}) 
takes the form
\bea
| \Psi \rangle& =&\sum _{\{(x_j,\s_j)\}}\Psi _{\s_1,\cdots,\s_N}
  (x_1,\cdots,x_N)c^\dagger
  _{x_1\s_1}\cdots c^\dagger_{x_N\s_N} | 0 \rangle,\no\\
\Psi_{\s_1,\cdots,\s_ N}(x_1,\cdots,x_N)
&=&\sum _P \e _P A_{\s_{Q1},\cdots,\s_{QN}}(k_{PQ1},\cdots,k_{PQN})
\exp (i\sum ^N_{j=1} k_{P_j}x_j),
\eea
where the summation is taken over all permutations and negations of
$k_1,\cdots,k_N,$ and $Q$ is the permutation of the $N$ particles such that
$1\leq   x_{Q1}\leq   \cdots  \leq  x_{QN}\leq   L$.
The symbol $\e_P$ is a sign factor $\pm1$ and changes its sign
under each 'mutation'. Substituting the wavefunction into  the
eigenvalue equation $ H| \Psi  \rangle = E | \Psi \rangle $,
one gets
\bea
A_{\cdots,\s_j,\s_i,\cdots}(\cdots,k_j,k_i,\cdots)&=&S_{ij}(
    k_i,k_j)
    A_{\cdots,\s_i,\s_j,\cdots}(\cdots,k_i,k_j,\cdots),\no\\
A_{\s_i,\cdots}(-k_j,\cdots)&=&s^L(k_j;p_{1\a_i})A_{\s_i,\cdots}
    (k_j,\cdots),\no\\
A_{\cdots,\s_i}(\cdots,-k_j)&=&s^R(k_j;p_{L\s_i})A_{\cdots,\s_i}(\cdots,k_j),
\eea
where $S_{ij}(k_i,k_j) $ are
the two-particle scattering matrices,
\bea
S_{ij}(k_i,k_j)^{11}_{11}&=&S_{ij}(k_i,k_j)^{22}_{22}=1,\no\\
S_{ij}(k_i,k_j)^{12}_{12}&=&S_{ij}(k_i,k_j)^{21}_{21}=
  \frac{\sin(\l_i-\l_j)}{\sin (\l_i-\l_j-i \g)},\no\\
S_{ij}(k_i,k_j)^{12}_{21}&=&
   e^{-i\;(\l_i-\l_j)}\frac{\sin i\g}{\sin(\l_i-\l_j-i\g)},\no\\
S_{ij}(k_i,k_j)^{21}_{12}&=&
   e^{i\;(\l_i-\l_j)}\frac{\sin i\g}{\sin(\l_i-\l_j-i\g)}
\eea
with $\l_j$ being suitable particle
rapidities related to the quasi-momenta $k_j$ of the electrons by
\cite{Bar95}
\bea
k(\l)&=&\lt\{
\begin{array}{l}
\Theta(\l,a),~~~~~~~~~~\e=+,\\
\pi-\Theta (\l,a),~~~~~\e=-
\end{array}
     \rt.\no\\
\Theta (\l,a)&=&2 \arctan (\coth a \tan \l),\no\\
a&=& \frac {1}{4} \lt\{ \ln [\frac {\sinh \frac {1}{2}(\eta +\g)}
 {\sinh \frac {1}{2}(\eta -\g)}]-\g \rt\},
\eea
and 
$s^L(k_j;p_{1\s_i})$,
 $~s^R(k_j;p_{L\s_i})$ are the boundary scatering matrices,
\bea
s^L(k_j;p_{1\s_i})&=&\frac {1-p_{1\s_i}e^{ik_j}}
{1-p_{1\s_i}e^{-ik_j}},\no\\
s^R(k_j;p_{L\s_i})&=&\frac {1-p_{L\s_i}e^{-ik_j}}
{1-p_{L\s_i}e^{ik_j}}e^{2ik_j(L+1)}
\eea
with
\bea
p_{1\a}&\equiv& p_1=-\mu ^{-1} +
  \frac {\mu -\mu ^{-1}}
  {2\sinh \frac {\g(2-\xi_-)}{2}}
   e^{-\g (1-\frac {\xi_-}{2})},\no\\
p_{L\a}&\equiv& p_L=-\mu+
  \frac {\mu -\mu ^{-1}}
  {2\sinh \frac {\g(2-\xi_+)}{2}}
   e^{\g (1-\frac {\xi_+}{2})}.
\eea
Then, the diagonalization of Hamiltonian (\ref{hamiltonian}) reduces 
to solving  the following matrix  eigenvalue equation
\beq
T_jt= t,~~~~~~~j=1,\cdots,N,
\eeq
where $t$ denotes an eigenvector on the space of the spin variables
and $T_j$ takes the form
\beq
T_j=S_j^-(k_j)s^L(-k_j;p_{1\s_j})R^-_j(k_j)R^+_j(k_j)
    s^R(k_j;p_{L\s_j})S^+_j(k_j)
\eeq
with
\bea
S_j^+(k_j)&=&S_{j,N}(k_j,k_N) \cdots S_{j,j+1}(k_j,k_{j+1}),\no\\
S^-_j(k_j)&=&S_{j,j-1}(k_j,k_{j-1})\cdots S_{j,1}(k_j,k_1),\no\\
R^-_j(k_j)&=&S_{1,j}(k_1,-k_j)\cdots S_{j-1,j}(k_{j-1},-k_j),\no\\
R^+_j(k_j)&=&S_{j+1,j}(k_{j+1},-k_j)\cdots S_{N,j}(k_N,-k_j).
\eea
This problem can  be solved using the algebraic Bethe ansatz method.
The Bethe ansatz equations are 
\bea
e^{ik_j2(L+1)}\zeta(k_j;p_1)\zeta(k_j;p_L)
&=&\prod ^{M}_{\a =1}\frac {\sin 
  (\l_j-\Lambda _{\a}+\frac {i\g}{2})}{\sin
  (\l_j-\Lambda _{\a} -\frac {i\g}{2})}\frac{\sin
  (\l_j+\Lambda _{\a}+\frac {i\g}{2})}
  {\sin (\l_j+\Lambda _{\a} -\frac {i\g}{2})},\no\\
\prod ^{N}_{j=1}
  \frac {\sin (\Lambda _{\a}-\l _j+\frac{i\g}{2})}
  {\sin (\Lambda _{\a}-\l _j-\frac{i\g}{2})}
  \frac {\sin (\Lambda _{\a}+\l _j+\frac{i\g}{2})}
   {\sin (\Lambda _{\a}+\l _j-\frac{i\g}{2})}
&=&
\prod ^{M}_{\stackrel {\b =1}{\b \neq \a}}
  \frac {\sin (\Lambda _{\a}-\Lambda _{\b}
   +i\g)}{\sin (\Lambda _{\a}-\Lambda _{\b}
  -i\g)}\frac {\sin (\Lambda _{\a}+\Lambda _{\b}
  +i\g)}{\sin (\Lambda _{\a}+\Lambda _{\b}-i\g)},
\eea
where $\zeta (k;p)= (1-pe^{-ik})/(1-pe^{ik})$.
The energy eigenvalue $E$ of the  model is given by
$E=-2\sum ^N_{j=1}\cos k_j$ (up to an unimportant additive constant,
which we have dropped).

\sect{Conclusion \label{con}}

We have proposed a very general graded RE algebra and developed the
corresponding boundary QISM.
In formulating our general formalism in section II
we do not impose any constraint
conditions on the quantum R-matrices, and  therefore our formalism
applies to all lattice boundary systems
where an invertible R-matrix exists. Two nontrivial
examples are the supersymmetric versions of
the models of two coupled and three coupled one-dimensional
$XY$ spin chains \cite{Bar91}, where the fermionic R-matrices do not possess
the so-called crossing-unitarity and  so no isomorphism between the boundary
K-matrices $K_-$ and $K_+$ exists. We have also considered the
important ``special" case in which the R-matrices are associated with
finite-dimensional representations of the quantum affine superalgebra
$U_q[{\cal G}^{(k)}]$ for generic $q$.
For such a case the two REs are isomorphic to each other since, as is
shown in the Appendix, all such R-matrices enjoy the crossing-unitarity
property.

We have applied our general formalism to study
integrable open-boundary conditions for the
$q$-deformed supersymmetric $U$ model of strongly correlated electrons.
The  quantum integrability
of the boundary system is established by the fact
that the corresponding Hamiltonian  may be embedded into
a one-parameter family of commuting transfer matrices. Moreover, the Bethe
ansatz equations are derived by means of the coordinate space Bethe ansatz
approach. This provides us with a basis for computing the finite-size
corrections to the low-lying energies in the system, which in turn allow
us to use the boundary conformal field theory technique to study
the critical properties of the boundary model.

As mentioned in section III, the Hamiltonian (\ref{hamiltonian}) 
becomes $U_q[gl(2|1)]$-invariant
in the limits $\xi _- \rightarrow -\infty$ and $\xi _+ \rightarrow  \infty$.
It is very interesting to see whether or not the Bethe
states constructed above constitute the highest weight states for 
$U_q[gl(2|1)]$. This property is crucial in understanding the completeness
of the Bethe states. A similar problem has been studied \cite{Gon94}
for the $U_q[gl(2|1)]$-invariant open $t-J$ chain. It seems
interesting to note that although the completeness problem for the
Bethe states are well studied in the periodic case for the Hubbard
model and the supersymmetric $t-J$ model \cite{Ess92}, it remains largely
unexplored in the non-periodic (or open) boundary case.
Another interesting question is to derive the Bethe ansatz
equations  using the algebraic Bethe ansatz approach.

\vskip.3in
%\acknowledgments
Y.-Z.Z and H.-Q.Z are supported by Australian Research Council, University of
Queensland New Staff Research Grant and External Support Enabling Grant. H.-Q.Z
would like to thank the Department of Mathematics, University of
Queensland, for kind hospitality.
He is also partially supported by the National Natural Science Foundation
of China and Sichuan Young Investigators Science and Technology Fund.

\appendix

\sect{On the crossing-unitarity}

So far in literature, the discussions on the crossing-unitarity of
R-matrices have been on a case by case basis and the crossing-unitarity
of a given R-matrix has been checked by brute force.

In this appendix, we show that 
all R-matrices associated with
finite-dimensional representations of the quantum affine superalgebra 
$U_q[{\cal G}^{(k)}]$ ($k=1,2$) for
generic $q$, where ${\cal G}$ is any simple Lie superalgebra,
enjoy the crossing unitarity property.

Let us first of all recall some facts about the affine superalgebra
${\cal G}^{(k)}$. Let ${\cal G}_0$ be the fixed point subalgebra under
the diagram automorphism $\hat{\tau}$ of ${\cal G}$ of order $k$. 
In the case of $k=1$, we have ${\cal G}_0\equiv{\cal G}$.
For $k=2$ we may decompose ${\cal G}$ as ${\cal G}_0$ plus a ${\cal
G}_0$-representation ${\cal G}_1$ of ${\cal G}$. Let
\beq
\psi=\lt\{
\begin{array}{l}
{\rm highest~ root~ of}~ {\cal G}_0\equiv {\cal G}, ~~{\rm for}~ k=1,\\
{\rm highest~ weight~ of~ the}~ {\cal G}_0-{\rm representation}~ 
  {\cal G}_1, ~~{\rm for}~ k=2.
\end{array}
\rt.
\eeq
Following the usual convention, we denote the weight
of ${\cal G}^{(k)}$ by $\L\equiv (\l,\kappa,\tau)$, where $\l$ is a
weight of ${\cal G}_0$. With this 
convention the nondegenerate form $(~,~)$ induced on the weights
can be expressed as
\beq
(\L,\L')=(\l,\l')+\kappa\tau'+\kappa'\tau.
\eeq
Let $\widehat{\rho}$ denote the distinguished dominant weight
of ${\cal G}^{(k)}$ satisfying $(\widehat{\rho},\a_i)=\frac{1}{2}(\a_i,
\a_i)$, where $\a_i~(0\leq i\leq r)$ are simple roots of ${\cal
G}^{(k)}$. Then $\widehat{\rho}$ is given by
\beq
\widehat{\rho}=\rho+g\g,~~~~~\g=(0,1,0),
\eeq
where $g=\frac{k}{2}(\psi,\psi+2\rho)$ and $\rho$ is 
the graded half sum of positive roots of ${\cal G}_0$. 

We shall not give the defining relations for $U_q[{\cal G}^{(k)}]$, 
but mention that the actions of coproduct and antipode on its 
generators $\{h_i,~e_i,~f_i,~d,~0\leq i\leq r\}$ are given by
\bea
\D(h_i)&=&h_i\otimes 1+1\otimes h_i,~~~~\D(d)=d\otimes 1+1\otimes
          d,\no\\
\D(e_i)&=&e_i\otimes q^{-\frac{h_i}{2}}+q^{\frac{h_i}{2}}\otimes
          e_i,~~~~
\D(f_i)=f_i\otimes q^{-\frac{h_i}{2}}+q^{\frac{h_i}{2}}\otimes
          f_i,\no\\
S(a)&=&-q^{-\widehat{\rho}}\;a\;q^{\widehat{\rho}},~~~~\forall 
       a=d,\;h_i,\;e_i,\;f_i.
\eea
Define an automorphism $D_z$ of $U_q[{\cal G}^{(k)}]$ by
\beq
D_z(e_i)=z^{\d_{i0}}e_i,~~~~~D_z(f_i)=z^{-\d_{i0}}f_i,~~~~
D_z(h_i)=h_i,~~~~D_z(d)=d.
\eeq
Then 
\beq
S^2(a)=q^{-2\rho}\,D_{q^{-\frac{2}{k}g}}\,(a)\,q^{2\rho},~~~~~
       S^{-2}(a)=q^{2\rho}\,D_{q^{\frac{2}{k}g}}\,(a)\,q^{-2\rho},~~~~
       \forall a\in U_q[{\cal G}^{(k)}],
\eeq
which can be checked on the generators [remembering that the simple
roots associated with $e_0,~f_0$ are $\a_0=\pm(\frac{1}{k}\d-\psi)$,
respectively, where $\d=(0,0,1)$].
We define the right dual module $V^*$ and
left dual module ${}^*V$ of $V$ by
\beq
\pi_{V^*}(a)=\pi_V(S(a))^{st},~~~~~~
  \pi_{{}^*V}(a)=\pi_V(S^{-1}(a))^{st},
\eeq
respectively. Here $st$ is the supertransposition operation defined by
\beq
(A_{ab})^{st}=(-1)^{[a]([a]+[b])}A_{ba}.
\eeq
Note that in general
$((A_{ab})^{st})^{st}=(-1)^{[a]+[b]}A_{ab}\neq A_{ab}$. Let $ist$ be the
inverse operation of $st$ such that $((A_{ab})^{st})^{ist}=
((A_{ab})^{ist})^{st}=A_{ab}$. Then
\beq
(A_{ab})^{ist}=(-1)^{[b]([a]+[b])}A_{ba}=(-1)^{[a]+[b]}(A_{ab})^{st},
   \label{ist-st}
\eeq
or $A^{ist}=\t A^{st} \t$, where $\t$ is a diagonal matrix with
elements $\t_{ab}=(-1)^{[a]}\d_{ab}$.

By arguments similar to those in \cite{Res90}, one can show that 
\beq
R^{V^*,W}(z)=(R^{VW}(z)^{-1})^{st_1},~~~~~
  R^{V,{}^*W}(z)=(R^{VW}(z)^{-1})^{st_2}.
\eeq
It follows from the representations for $R^{V^{**},W}(z)$ and
$R^{V,{}^{**}W}(z)$ that for
any pair of finite dimensional $U_q[{\cal G}^{(k)}]$-modules $V$ and
$W$, the R-matrix satisfies the following crossing-unitarity relations
\bea
(((R^{VW}(z)^{-1})^{st_1})^{-1})^{st_1}&=&(\pi_V(q^{-2\rho})\otimes 1_W)
    ((R^{VW}(zq^{-\frac{2}{k}g}))^{st_1})^{st_1}(\pi_V(q^{2\rho})
    \otimes 1_W),\no\\
(((R^{VW}(z)^{-1})^{st_2})^{-1})^{st_2}&=&(1_V\otimes \pi_W(q^{2\rho}))
    ((R^{VW}(zq^{\frac{2}{k}g}))^{st_2})^{st_2}(1_V\otimes\pi_W(q^{-2\rho})).
    \label{cu-twisted}
\eea
Note also that
\beq
(\pi_V(q^{\pm 2\rho})\otimes\pi_W(q^{\pm 2\rho}))R^{VW}(z)
  =R^{VW}(z)(\pi_V(q^{\pm 2\rho})\otimes\pi_W(q^{\pm 2\rho})).\label{rho}
\eeq

Let us remark that if one uses the opposite coproduct and antipode of
$U_q[{\cal G}^{(k)}]$,
\bea
\bar{\D}(h_i)&=&h_i\otimes 1+1\otimes h_i,~~~~\bar{\D}(d)=d\otimes 1+1\otimes
          d,\no\\
\bar{\D}(e_i)&=&e_i\otimes q^{\frac{h_i}{2}}+q^{-\frac{h_i}{2}}\otimes
          e_i,~~~~
\bar{\D}(f_i)=f_i\otimes q^{\frac{h_i}{2}}+q^{-\frac{h_i}{2}}\otimes
          f_i,\no\\
\bar{S}(a)&=&-q^{\widehat{\rho}}\;a\;q^{-\widehat{\rho}},~~~~\forall 
       a=d,\;h_i,\;e_i,\;f_i,
\eea
and denote the corresponding R-matrix by $\bar{R}(z)$, then the similar
arguments as above give rise to the following crossing-untarity
relations:
\bea
(((\bar{R}^{VW}(z)^{-1})^{st_1})^{-1})^{st_1}&=&(\pi_V(q^{2\rho})\otimes 1_W)
    ((\bar{R}^{VW}(zq^{\frac{2}{k}g}))^{st_1})^{st_1}(\pi_V(q^{-2\rho})
    \otimes 1_W),\no\\
(((\bar{R}^{VW}(z)^{-1})^{st_2})^{-1})^{st_2}&=&(1_V\otimes \pi_W(q^{-2\rho}))
    ((\bar{R}^{VW}(zq^{-\frac{2}{k}g}))^{st_2})^{st_2}
    (1_V\otimes\pi_W(q^{2\rho})).\label{cu-twisted-bar}
\eea

%\newpage
%\vskip.3in

\end{document}